\documentclass[journal=apchd5,manuscript=article]{achemso}

\usepackage{amsmath}
\usepackage{natbib}
\usepackage{graphicx}
\usepackage[usenames,dvipsnames]{color}
\pdfoutput=1

\usepackage{amssymb}

\usepackage{color}
\usepackage{url}

\renewcommand{\eqref}[1]{\mbox{Eq.~(\ref{#1})}}

\usepackage[version=3]{mhchem} 
\usepackage[T1]{fontenc}       

\author{Claudia Triolo}
\author{Adriano Cacciola}
\author{Salvatore Patan\`{e}}
\author{Rosalba Saija}
\author{Salvatore Savasta} \email{ssavasta@unime.it}
\affiliation[Messina]{Dipartimento di Scienze Matematiche e Informatiche, Scienze Fisiche e Scienze della Terra (MIFT), Universit\`{a} di Messina, I-98166 Messina, Italy}
\alsoaffiliation[Riken]{CEMS, RIKEN, Saitama 351-0198, Japan}
\author{Franco Nori}
\affiliation[Riken]{CEMS, RIKEN, Saitama 351-0198, Japan}
\alsoaffiliation{Physics Department, The University of Michigan, Ann Arbor, Michigan 48109-1040, USA}

\title{Spin-Momentum Locking\\* in the Near Field of Metal Nanoparticles}


\begin{document}

\begin{tocentry}
\begin{center}
	\includegraphics[scale=0.21]{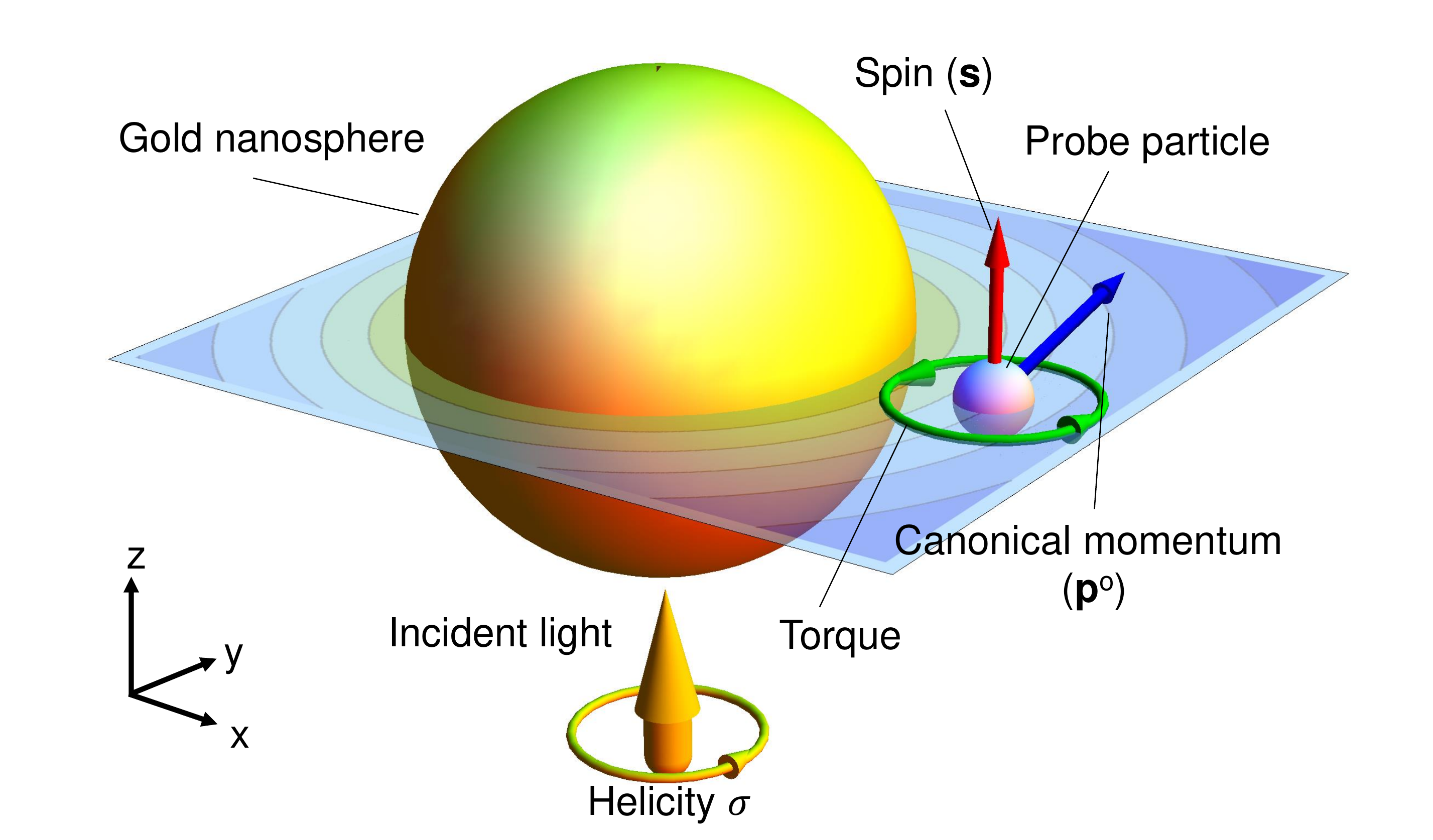}
\end{center}
\vspace{1 cm}
	{\bf For Table of Contents Use Only}. 
	
	{\bf Spin-Momentum Locking in the Near Field of Metal Nanoparticles}, C. Triolo, A. Cacciola, S. Patan\`{e}, R. Saija, S. Savasta. F. Nori.
	\vspace{0.2 cm}
	
	Schematic representation of the configuration used for the scattering calculations. Under the resonance condition, the exciting incident wave induces a strong enhancement of the electromagnetic field around the sphere, that rapidly decays  from the particle surface and produces interesting effects related to the orbital and spin momenta of light and their SOI. The force and the torque produced by the orbital and spin momentum, respectively, acan be investigated by considering a probe-particle near the gold nanosphere.	
\end{tocentry}

\begin{abstract}
Light carries both spin and  momentum. Spin-orbit interactions of light come into play at the subwavelength scale of nano-optics and nano-photonics, where they determine the behaviour of light. These phenomena, in which the spin  affects and controls the spatial degrees of freedom of light, are attracting rapidly growing interest. 
Here we present results on the spin-momentum locking in the \textit{near field} of metal nanostructures supporting localized surface resonances.  These systems can confine light to very small dimensions below the diffraction limit, leading to a striking near-field enhancement. In contrast to the \textit{propagating} evanescent waves of surface plasmon-polariton modes, the electromagnetic near-field of localized surface resonances does not exhibit a definite position-independent  momentum or polarization. Close to the particle, the canonical momentum is almost tangential to the particle surface and rotates when moving along the surface. The direction of this rotation can be controlled by the spin of the incident light.


\end{abstract}

\hspace{1 cm}


According to Maxwell's theory of electromagnetism, an electromagnetic wave carries both momentum and angular momentum (AM), which can be transferred to a reflecting or absorbing surface hit by the wave \cite{Molina2007,loudon2012}.
The simplest example of an optical field carrying momentum and spin angular momentum is an elliptically-polarized plane wave. Assuming the free space propagation along the $z$-axis, the complex electric field of this wave can be written as,
\begin{equation}\label{planewave1}
{\bf E} ({\bf r}) = A \left( \bar {\bf x}\, \cos \frac{\theta}{2} + \bar {\bf y} \sin \frac{\theta}{2}\,  e^{i \phi}  \right) e^{i k z}\, ,
\end{equation}
where $A$ is the wave amplitude, $\bar {\bf x}$ and $\bar {\bf y}$ are unit vectors, $k = \omega/c$ is the wave number, and the angles $\theta$ and $\phi$ determine the polarization state. Throughout the paper we imply monochromatic fields, omitting the time-evolution factor $e^{-i \omega t}$.

The momentum ${\bf p}$ and spin AM ${\bf s}$ densities of the wave described by  Eq.\ (\ref{planewave1}) are longitudinal:
\begin{equation}\label{planewave2}
{\bf p} = \frac{w}{\omega} k\, \bar {\bf z}\, , \hspace{1 cm} {\bf s} = \frac{w}{\omega}\, \sigma\, \bar {\bf z}\, ,
\end{equation}
where $w = \gamma \omega  A^2$ is the energy density [$\gamma = (8 \pi \omega)^{-1}$ in Gaussian units], and $\sigma = \sin \theta\, \sin \phi \in [-1,1]$ is the helicity parameter.
The momentum ($\propto k$) describes the propagation of the wave, while the spin AM ($\propto \sigma$) characterizes the independent polarization degree of freedom.

Real optical beams can differ significantly from the idealized plane wave described in Eq.\ (\ref{planewave1}). However, traditional macroscopic optics can maintain this picture, still treating the spatial and polarization properties of light as independent. For example, the first can be manipulated by lenses or prisms, while the latter can be independently affected by polarizers or waveplates. At the subwavelength scales of nano-optics, photonics and plasmonics, however, spin and orbital properties become strongly coupled with each other. 

The spin-orbit interactions (SOI) of light are nowadays a rapidly growing area of research, which is of both fundamental and practical interest \cite{bliokh2015, bliokh2015spin, Antognozzi2016}. These studies reveal interesting connections between optical SOI and fundamental quantum mechanics or field-theory problems involving optical momentum and spin. Moreover, the miniaturization of optical devices and the fast development of nano-photonics require to consider the SOI of light. Indeed, it turns out that most optical processes (e.g., propagation, reflection, focusing, scattering, and diffraction) are strongly influenced by the SOI at subwavelength scales \cite{petersen2014chiral,Donato2014,o2014spin, Zhang2017}.

Spin-dependent perturbations of the light trajectory, which is a manifestation of the spin-Hall effect, in a gradient-index medium is a first important example of SOI \cite{bliokh2015quantum,bliokh2015}.
Optical spin-momentum locking was recently observed in many experiments exploiting evanescent waves. For example, coupling incident circularly-polarized light to the evanescent tails of surface or waveguide modes, results in a strong spin-controlled unidirectional excitation of these modes. This is a direct manifestation of the extraordinary transverse spin of evanescent waves related to the quantum spin-Hall effect of light. It has also been shown that the focusing of circularly polarized light by a high-numerical-aperture lens, or the scattering by a small particle, generates a spin-dependent optical vortex in the output field.

Here we investigate optical SOI in the near-field region of metallic nanoparticles.
When light interacts with metal nanoparticles and nanostructures, it can excite collective oscillations, known as localized surface plasmons (LSPs), which can confine light to very small dimensions below the diffraction limit \cite{giannini2011plasmonic,triolo2015near}.
The angular spectrum representation shows that radiation re-emitted by a localized source is a combination of travelling and evanescent waves \cite{moreno2013}. The latter largely dominate the near-field region around metallic nanoparticles supporting LSPs. In contrast to the surface plasmon-polariton modes, the near field of LSP resonances does not exhibit a definite position-independent  momentum or polarization.
Very recently, the concept of local angular momentum  as a figure of merit for the
design of nanostructures that provide large field gradients has been proposed \cite{Alabastri2016}. These systems
offer the opportunity to investigate spin-momentum locking and more general SOI of light for complex multimode evanescent fields.
The results presented here show that spin-momentum locking, spin controlled unidirectional propagation of light, and spin-controlled optical forces can also be observed in the \textit{near field} of metal nanoparticles.

\section*{Results}

For a vector field, the momentum of light is usually defined by the Poynting vector ${\bf p}$  which, in the simplest case of a homogeneous plane electromagnetic wave,  is aligned with the wavevector $\bf k$ [see Eq. (\ref{planewave2})]. However, in more complicated (yet typical cases of) structured optical fields (such as optical vortices and near-field phenomena), the direction of the Poynting vector can differ from the wavevector direction \cite{bekshaev2015transverse,rodriguez2013near,bliokh2012spatiotemporal,shitrit2011optical}. In these cases,  the Poynting vector $\bf p$ acquires an additional spin momentum density $\bf p^{\rm s}$,  introduced for the first time by Belifante \cite{belinfante1940,ohanian1986spin}, and can be expressed \cite{bliokh2015} as a sum of canonical and spin contributions: ${\bf p}= {\bf p}^{\rm o} +{\bf p}^s$. 

In terms of the electric ${\bf E}$ and magnetic {\bf H} components of the optical field, we have \cite{Bliokh2014}:
\begin{equation}\label{planewave3}
{\bf p}^{\rm o} = \frac{\gamma}{2} {\rm Im}[\bf E^\ast \cdot (\bf \nabla) \bf E+ \bf H^\ast \cdot (\bf \nabla) \bf H]\,
\end{equation}
\begin{equation}\label{planewave4}
{\bf s}= \frac{\gamma}{2} {\rm Im} [\bf E^\ast \times \bf E + \bf H^\ast \times \bf H]\, , \hspace{1 cm}
{\bf p}^{\rm s}= \frac{1}{2} \bf \nabla \times \bf s \, .
\end{equation}
The optical momentum and spin densities can be measured experimentally by placing a small absorbing particle in the field and observing its linear $(\bf F \propto \bf p)$ and spinning $(\bf T \propto \bf s)$ motion \cite{Bliokh2014,adachi2007orbital,Antognozzi2016,rodriguez2015lateral}. This description is also valid for the canonical and the spin momenta of evanescent waves. Considering the total internal reflection of a polarized plane wave at the glass-air interface, the canonical momentum density in the evanescent field in air is proportional to its longitudinal wavevector ${\bf p}^{\rm o} \propto k_z \bf \bar{z}$ (where $\bf \bar{z}$ indicates the propagation direction). However, at the same time, the Poynting vector has an unusual transverse component, which depends on the spin \cite{Bliokh2014}:
\begin{equation} \label{planewave5}
{\bf s} = \frac{\tilde w}{\omega}\left(\sigma  \frac{k}{k_z} {\bf \bar{z}} + \frac{\kappa}{k_z} {\bf \bar{y}}\right)\, , \hspace{1 cm}
{\bf p}^{\rm s} = \frac{\tilde w}{\omega}\left(- \frac{\kappa^2}{k_z} {\bf \bar{z}} + \sigma \frac{\kappa k}{k_z} {\bf \bar{y}}\right)
\end{equation}
where $\tilde {w}=\gamma \omega |A|^2 e^{-2\kappa x}$, $k_z$ is the longitudinal wavenumber and $\kappa = \sqrt{k_z^2 - k^2}$ is the exponential decay rate. The second term in each of the two Eqs. in (\ref{planewave5}), provides a transverse component to the spin and momentum. These are specific features of the evanescent fields. The transverse component of the momentum becomes proportional to the helicity $\sigma$, while that acquired by the spin turns out to be helicity independent \cite{Bliokh2014}. Considering a dipole Rayleigh particle with equal electric and magnetic polarizabilities $\alpha = \alpha_{\rm e} = \alpha_{\rm m}$, the radiation pressure force on it is \cite{Bliokh2014} ${\bf F} = \gamma^{-1} {\rm Im}(\alpha) {\bf p}^o$. The resulting radiation pressure (longitudinal) force ``per photon'' is therefore $8 \pi \hbar \omega {\rm Im}(\alpha) k_z$. Since $k_z$ can exceed $k$ for evanescent waves, the force from the evanescence field can be higher than the force from a plane wave with the same local wave vector ${\bf k}$.

If we compare the radiation force that acts on a probe particle generated by a propagating wave or by an evanescent wave, in the latter this force will be larger than $k$ per photon. It turns out that, for ideal dipole Rayleigh particles,  ${\bf p}^{\rm s}$ does not contribute to the force exerted on it by the field. However, for larger or anisotropic probe particles, its value can be different from zero. The spin produces two radiation torque components on a probe particle. The longitudinal torque depends on the spin state, while the transverse torque is $\sigma$-independent and it occurs even for linearly-polarized incident light \cite{Bliokh2014}. 

  \begin{figure}[!ht]
  	\centering
  	\includegraphics[scale=0.33]{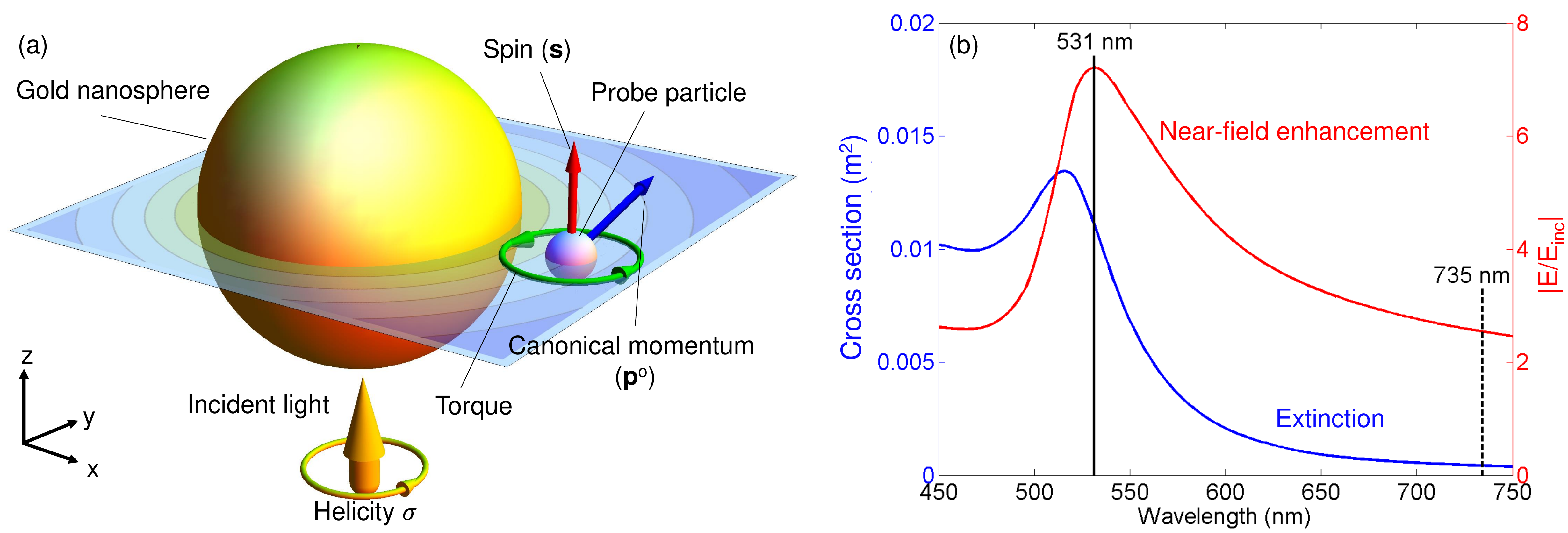}
  	\caption{(a) Schematic representation of the configuration used for the scattering calculations. The incident field propagates along the $z$-axis. Under the resonance condition, the exciting incident wave induces a strong enhancement of the electromagnetic field around the sphere, that rapidly decays  from the particle surface and produces interesting effects related to the orbital and spin momenta of light and their SOI. The force and torque produced by the orbital and spin momentum, respectively, are calculated by considering a probe-particle near the gold nanosphere. (b) Near-field enhancement $|E/E_{\rm inc}|^2$ (red curve) calculated on the equatorial plane, at  a distance $d=4$ nm from the particle surface, and the extinction efficiency cross-section spectra (blue curve) for a gold spherical nanoparticle (radius $a=40$ nm), calculated beyond the quasistatic approximation, by employing the Mie theory implemented within the T-matrix formalism \cite {borghese2007scattering,borghese2013superposition}. The resonance condition occurs in correspondence of the maximum near-field enhancement of the optical field at $\lambda=531$ nm (black vertical line). The dotted vertical line indicates an out-of-resonance wavelength ($\lambda=735$ nm), where we also calculated the canonical and spin momenta.    
  		\label{fig:1}}
  \end{figure}
    
Here we propose to exploit the near-field enhancement of LSPs resonances in order to investigate the orbital and spin momenta of light and their SOI in the near-field region of metallic nanoparticles.

\newpage
\subsection*{Circularly-polarized incident field}

Figure~1a shows a schematic representation of the configuration used here for the scattering calculation. The simplest possible geometry considered here involves only an incident propagating plane wave with amplitude $E_{\rm inc} = A$ [see Eq.~\ref{planewave1}] and a metallic sphere. 
Below, we analyze the characteristics of the scattered field ${\bf E}_{\rm sc}$ and of the total field, ${\bf E} = {\bf E}_{\rm inc} + {\bf E}_{\rm sc}$, in conjuction with the incident field characteristics ${\bf p}_{\rm inc}$, $w_{\rm inc}$, and ${\bf s}_{\rm inc}$, determined by Eq.~(\ref{planewave2}). Conventionally, the quantities related with the incident (scattered) field are marked by the subscripts ``inc'' (``sc''), and the total field characteristics are shown without subscript. For convenience, the momentum densities are normalized by the incident field momentum density, e.g., 
${\bf p}^o \to {\bf p}_{\rm n}^o = {\bf p}^o  / |{\bf p}_{\rm inc}|$, and the spin density is normalized by the incident field energy density according to  ${\bf s} \to {\bf s}_{\bf n} = (\omega / w_{\rm inc}) {\bf s}$.

Let the incident field be a circularly-polarized plane wave travelling along the $z$-axis. Under the resonance condition, the exciting wave induces a strong enhancement of the electromagnetic field around the sphere. We calculate the orbital momentum and the spin on the equatorial plane that goes through the center of the sphere and normal to the propagation direction. We consider a gold spherical nanoparticle of radius ($a=40$ nm) smaller than the effective wavelength $\lambda/\sqrt{\varepsilon_d}$, where $\varepsilon_d$ is the dielectric constant of the surrounding medium (through this work we use $\varepsilon_d =1$). Figure~1b displays the near-field enhancement $|E/E_{\rm inc}|^2$ (red curve),  and the extinction efficiency (blue curve) for a gold spherical nanoparticle.


Figure~2a shows the orbital momentum enhancement $|{\bf p}^{\rm o}|/|{\bf p_{\rm inc}}|$ and the module of the Poynting vector $|{\bf p}|/|{\bf p_{\rm inc}}|$ as a function of the distance $d$ from the nanoparticle surface. Close to the particle surface, due to the strong confinement of the scattered field, the Poynting vector acquires an additional component that depends on the spin, as defined in Eq. (\ref{planewave4}). This additional component, so-called \textit{spin momentum} ${\bf p}^{\rm s}$, produces a ``supermomentum" effect that causes the enhancement of $|{\bf p}^{\rm o}|$ (almost one order of magnitude greater than $|{\bf p}|$).
A similar feature characterizes also the spin density (calculated, but not shown). Figure~2b displays the amplitude of the normalized spin density ${\bf s}_{\rm n} = (\omega/ w_{\rm inc}) {\bf s}$ as a function of $d$. This ratio can vary between $-1$ and 1. For the scattering contribution, the curve decays approximately linearly with increasing $d$. This shows that the spin density associated with the scattered field decays with the distance more rapidly than the energy density. The spin density of the total field is significantly smaller than that of the scattered field for $d \lesssim 20$ nm. This lower spin density of the total field is caused by the interference between the incident and the scattered field that produces a spin reduction in the local field. The values of ${\rm s}_{n}$ for the scattered field indicate an intermediate spin state  between $\sigma = \pm 1$ (purely circular polarization) and $\sigma = 0$ (linear polarization), describing an elliptical polarization. However, the incident field is characterized by an exact spin state ($\sigma= \pm 1$). It results that the spin direction of the scattered field is almost opposite to that of the incident field. Owing to the rapid decay of the scattered field, moving away from the particle surface, spin cancellation rapidly increases, giving rise to a strong lowering of the spin of the total field. Increasing even more the distance $d$, the scattered field becomes negligible, the incident field prevails, and the spin value increases approaching 1. This explains the minimum value observed in Fig.~2b. 

The presence of the transverse component of the spin can be demonstrated considering the angle $\theta$ between the ${\bf p}^{\rm o}$ and $\bf s$ vectors (see Fig.~2c). In contrast to the longitudinal spin of the incident wave, the spin of the scattered field turns out to be almost completely transverse to the canonical momentum ${\bf p}^{\rm o}$, independently on the helicity ($\sigma=\pm 1$) of the incident light (see dotted curves in Fig.~2c). We observe, however, that the angle is slightly larger than $90^{○\circ} $ for $\sigma= -1$ and smaller for $\sigma=1$. 

\begin{figure}[!ht]
	\centering
	\includegraphics[width = \linewidth]{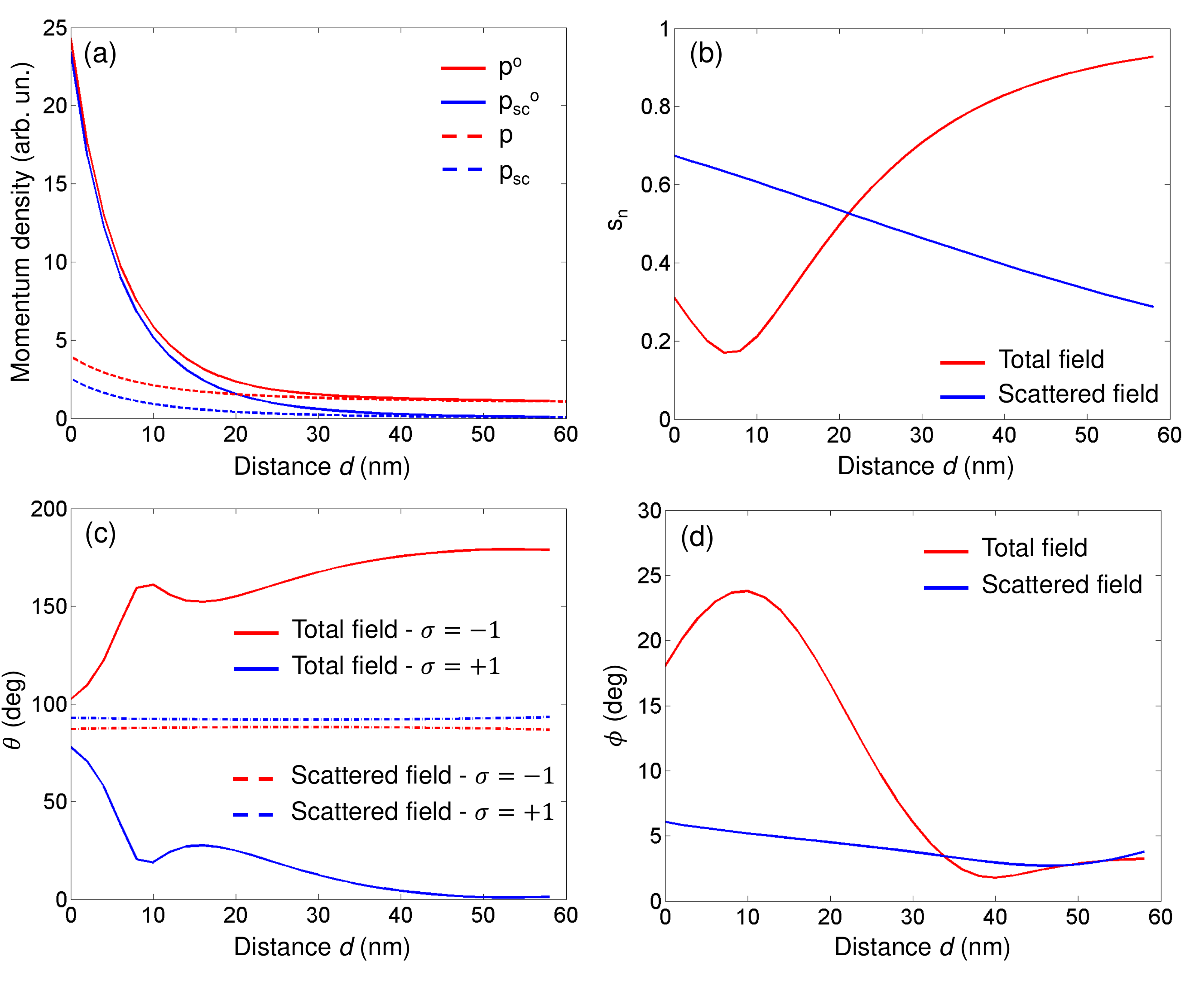}
	\caption{Circularly-polarized incident field at a wavelength of $\lambda=531$ nm (resonance condition): (a) Orbital momentum density enhancement $|\bf p^{\rm o}|/|\bf p_{\rm inc}|$ and Poynting vector enhancement $|\bf p|/|\bf p_{\rm inc}|$  as a function of the distance $d$ from the surface of the nanoparticle. This panel also shows these quantities obtained considering only the scattered field contribution: $|\bf p^{\rm o}_{\rm sc}|/|\bf p_{\rm inc}|$ and  $|\bf p_{\rm sc}|/|\bf p_{\rm inc}|$. (b) Modulus of the normalized spin density $\bf s_{\rm n}$ as a function of $d$ for the total and the scattered field. (c) Angle $\theta$ between $\bf p^{\rm o}$ and $\bf s$ as a function of distance $d$ for $\sigma = \pm 1$ incident polarization. (d) Angle $\phi$ between ${\bf p}$ and $\bf p^{\rm o}$.  The angles in panels (c) and (d) are also displayed considering the scattered contribution only. All the curves in  panels (a), (b), and (d) are equal for the two circular polarizations $\sigma = \pm 1$.
	All the displayed curves have been calculated for a gold sphere at the equatorial plane $xy$. 
		\label{fig:2}}
\end{figure}
\newpage
\begin{figure}[!ht]
	\centering
	\includegraphics[width = 14 cm]{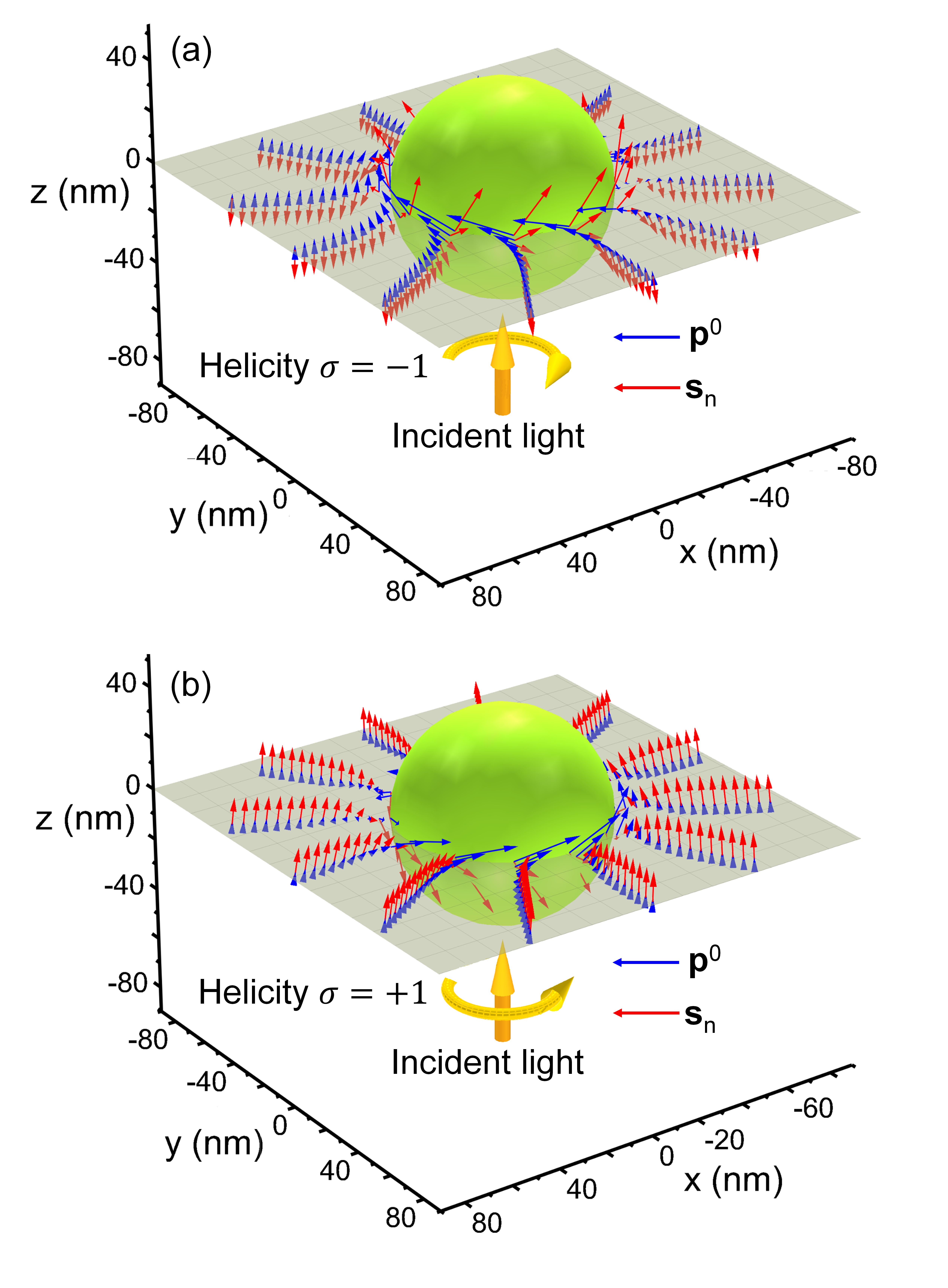}
	\caption{Vectors $\bf p^{\rm o}$ and ${\bf s}_{\rm n}$ displayed on the equatorial plane of the sphere, calculated at the plasmonic \textit{resonance} $\lambda = 531$ nm. (a) for  $\sigma = -1$ incident light  and (b) for  $\sigma = +1$. The origin corresponds to the centre of the nanosphere. Note that $\bf p^{\rm o}$ and ${\bf s}_{\rm n}$ close to the surface are nearly perpendicular.
		\label{fig:3}}
\end{figure}
\newpage

Figure~2c shows also the angle between ${\bf p}^{\rm o}$ and ${\bf s}$ for the total field. We notice that at the particle surface, where the scattered field largely dominates, the transverse component of the spin prevails and $\theta \approx 80^{\circ}$ (for $\sigma=+1$) and $\theta \approx 100^{\circ}$ (for $\sigma= - 1$). As expected, at increasing distances, the two angles (for $\sigma= \pm 1$) tend towards those describing the (longitudinal) spin direction of the incident waves:  $\theta_+ = 0^{\circ}$, and $\theta_- = 180^{\circ}$. We observe that the dependence of the angles on the distance is not monotonous. A local minimum (maximum) can be observed around $d \approx 10$ nm. It originates from the same cancellation effect determining the minimum in Fig.~2b.
Figure 2d displays the angle $\phi$ between $\bf p$ and ${\bf p}^{\rm o}$ as a function of $d$ and calculated for the scattered and total fields. 
We observe that, for the scattered fields, $\phi$ decays approximately linearly with increasing $d$, as the spin of the scattered field (see Fig.~2b). This behaviour can be understood noticing that the spin momentum is defined as ${\bf \nabla \times s}$.
Note that the values of $\phi$  are very small (its maximum value is about $6^{\circ}$) and indicate that the Poynting vector is almost coincident with the canonical momentum. The behavior of the $\phi$ angle is more complex for the total field, due to the interference between the incident and scattered fields. Indeed, because within $30$ nm from the particle surface the spin undergoes considerable variations (see red curve in Fig.~2b), in this region ${\bf p}^{\rm s}$ is larger, giving rise to a larger difference between the two vectors. The maximum value reached by the angle $\phi$ between them is almost $24^{\circ}$. This value is comparable to that calculated for an evanescent wave generated by a polarized propagating wave that undergoes total internal reflection at the glass-air interface. Considering an incidence angle of $45^{\circ}$ and a refractive index of glass $n=1.5$, the angle between $\bf p$ and ${\bf p}^{\rm o}$ is $\phi \sim 20^{\circ}$. 

The results described above, obtained by resonantly exciting the LSPs of a gold nanosphere, have shown that the optical field in the near-field region possesses remarkable properties related to the SOI. 
These interactions induce: (\textit{i}) the rise of a transverse spin which close to the particle surface is dominant; (\textit{ii}) the appearence of an extraordinary spin-dependent momentum, so that, in the near field, the canonical momentum  ${\bf p}^{\rm o}$ differs significantly from the Poynting vector $\bf p$, as in the case of propagating surface waves \cite{bliokh2015,Bliokh2014}. 

Recently, several experiments and numerical simulations have demonstrated notable spin-controlled unidirectional coupling between circularly-polarized incident light and transversely propagating surface or waveguide modes, which can be associated with the quantum spin-Hall effect of light  \cite{bliokh2015, bliokh2015spin,bliokh2015quantum}. Now we investigate this SOI effect in the near-field of a metallic nanoparticle. Figures~3a and 3b, displaying the directions of the canonical momentum and of the spin on the equatorial plane of the particle, clearly show that the spin of the incident light is able to control the direction of the canonical momentum ${\bf p}^{\rm o}$. Very close to the particle surface, the canonical momentum lies almost completely on the equatorial plane, with a small tilt along the propagation direction $\bar {\bf z}$ of the exciting field independent on the incident polarization. The position-dependent ${\bf p}^{\rm o}$ wraps around the sphere with the same clockwise or counter-clockwise rotation of the incident polarization. Hence, the helicity ($\sigma = \pm 1$) of the incident light determines the rotation direction of the canonical momentum  near the surface of the particle. Note that, close to the particle surface, the spin is almost opposite to the spin direction of the incident field and forms an angle of  $ \approx 90^{\circ}$ with the canonical momentum. As shown in Fig.~2c, the angle is larger (smaller) than $90^{\circ}$ for $\sigma = -1$ ($+1$). This difference is due to the spin-independent tilt of the momentum along the incident direction, mainly due to the contribution of the incident field to the total field. This contribution, owing to the rapid decay of the scattered field, becomes more relevant with the distance from the particle surface and it significantly affects both canonical momentum and spin. Specially, ${\bf p}^{\rm o}$ tends to align with the Poynting vector of the total field, and ${\bf s}$ varies much rapidly within 30 nm from the particle surface, rotating through the equatorial plane $xy$ of the nanosphere, and finally reaching the same longitudinal direction of the spin of the incident light. These effects represent a confirmation of the spin-orbit coupling \cite{bliokh2015spin} in the near-field region around a metallic nanoparticle which, due to the LSPs resonance, is dominated by the evanescent field.

\begin{figure}[!ht]
	\centering
	\includegraphics[width = \linewidth]{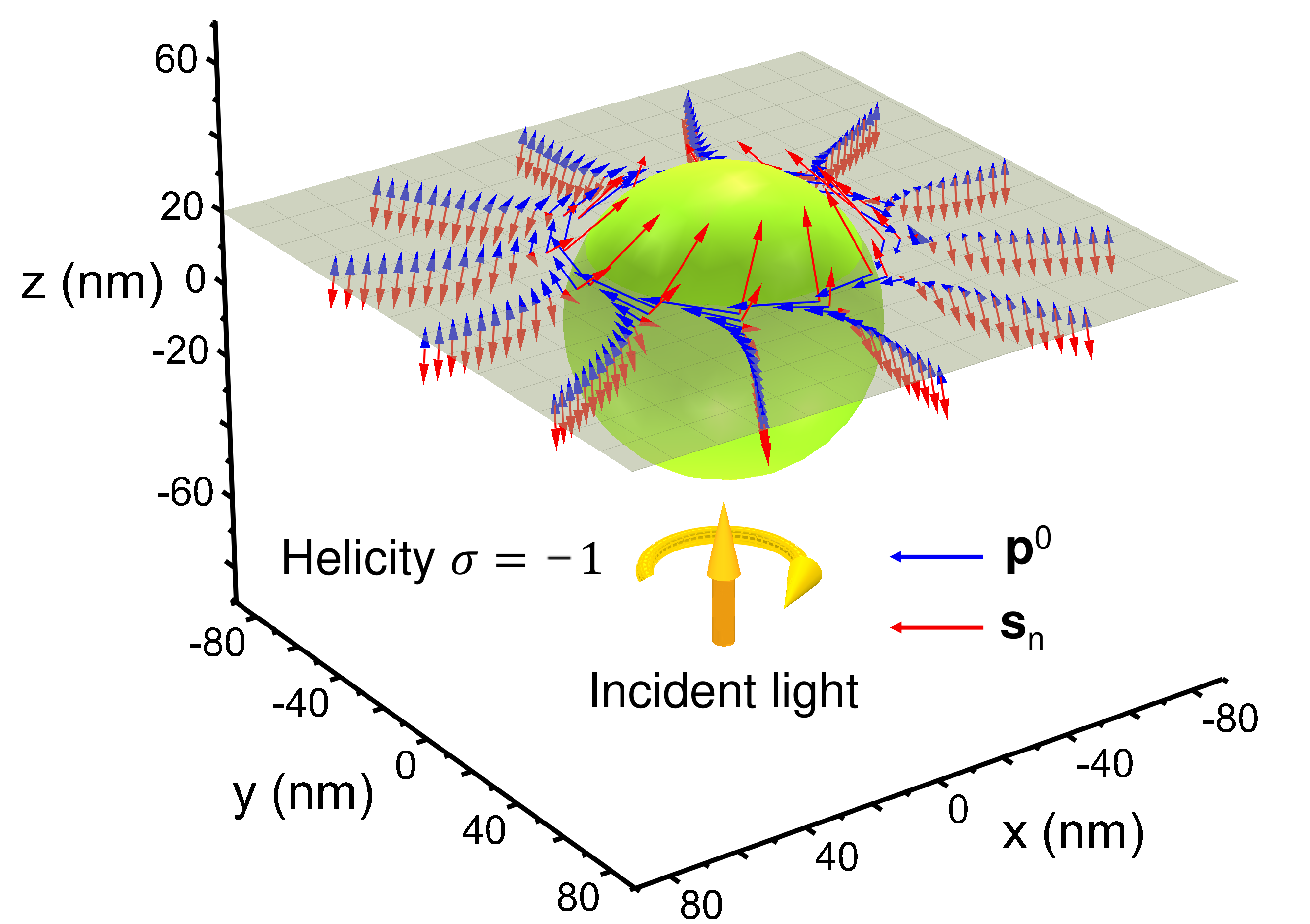}
	\caption{Vectors $\bf p^{\rm o}$ and ${\bf s}_{\rm n}$ displayed on a plane 20 nm above the equatorial plane of the sphere, calculated at the plasmonic \textit{resonance} $\lambda = 531$ nm, for  $\sigma = -1$ incident light.
		\label{fig:4}}
\end{figure}

\begin{figure}[!ht]
	\centering
	\includegraphics[width = \linewidth]{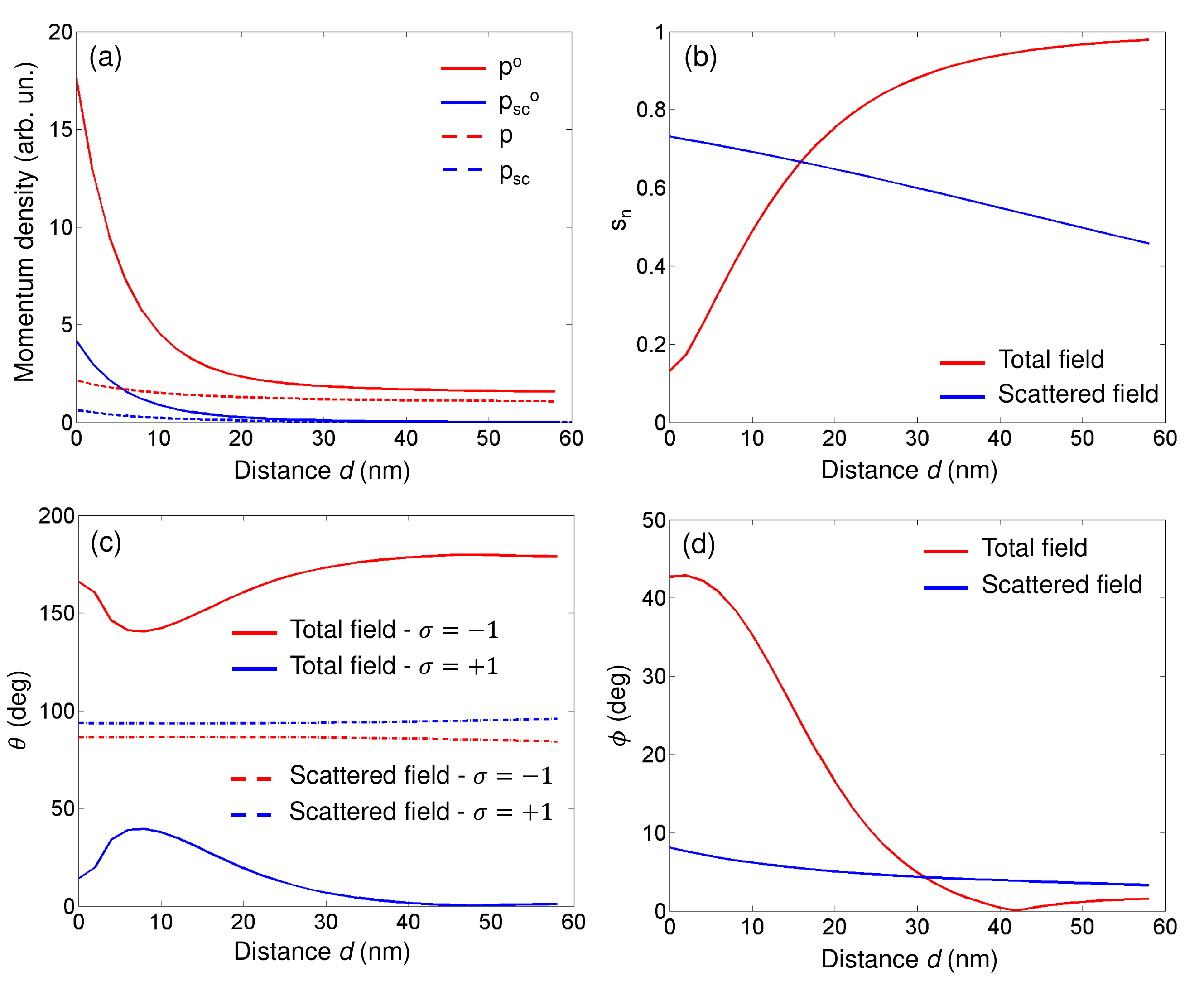}
	\caption{Circularly-polarized incident field at a wavelength of $\lambda=735$ nm: (a) Orbital momentum density enhancement $|\bf p^{\rm o}|/|\bf p_{\rm inc}|$ and Poynting vector enhancement $|\bf p|/|\bf p_{\rm inc}|$  as a function of the distance $d$ from the surface of the nanoparticle. This panel also shows these quantities obtained considering only the scattered field contribution: $|\bf p^{\rm o}_{\rm sc}|/|\bf p_{\rm inc}|$ and  $|\bf p_{\rm sc}|/|\bf p_{\rm inc}|$. (b) Modulus of the normalized spin density $\bf s_{\rm n}$ as a function of $d$ for the total and the scattered field. (c) Angle $\theta$ between $\bf p^{\rm o}$ and $\bf s$ as a function of distance $d$ for $\sigma = \pm 1$ incident polarization. (d) Angle $\phi$ between ${\bf p}$ and $\bf p^{\rm o}$. The angles in panels (c) and (d) are also displayed considering the scattered contribution only. All the curves in  panels (a), (b), and (d) are equal for the two circular polarizations $\sigma = \pm 1$.
	All the displayed curves have been calculated for a gold sphere at the equatorial plane $xy$. 
		\label{fig:5}}
\end{figure}

\begin{figure}[!ht]
	\centering
	\includegraphics[width = 14.5 cm]{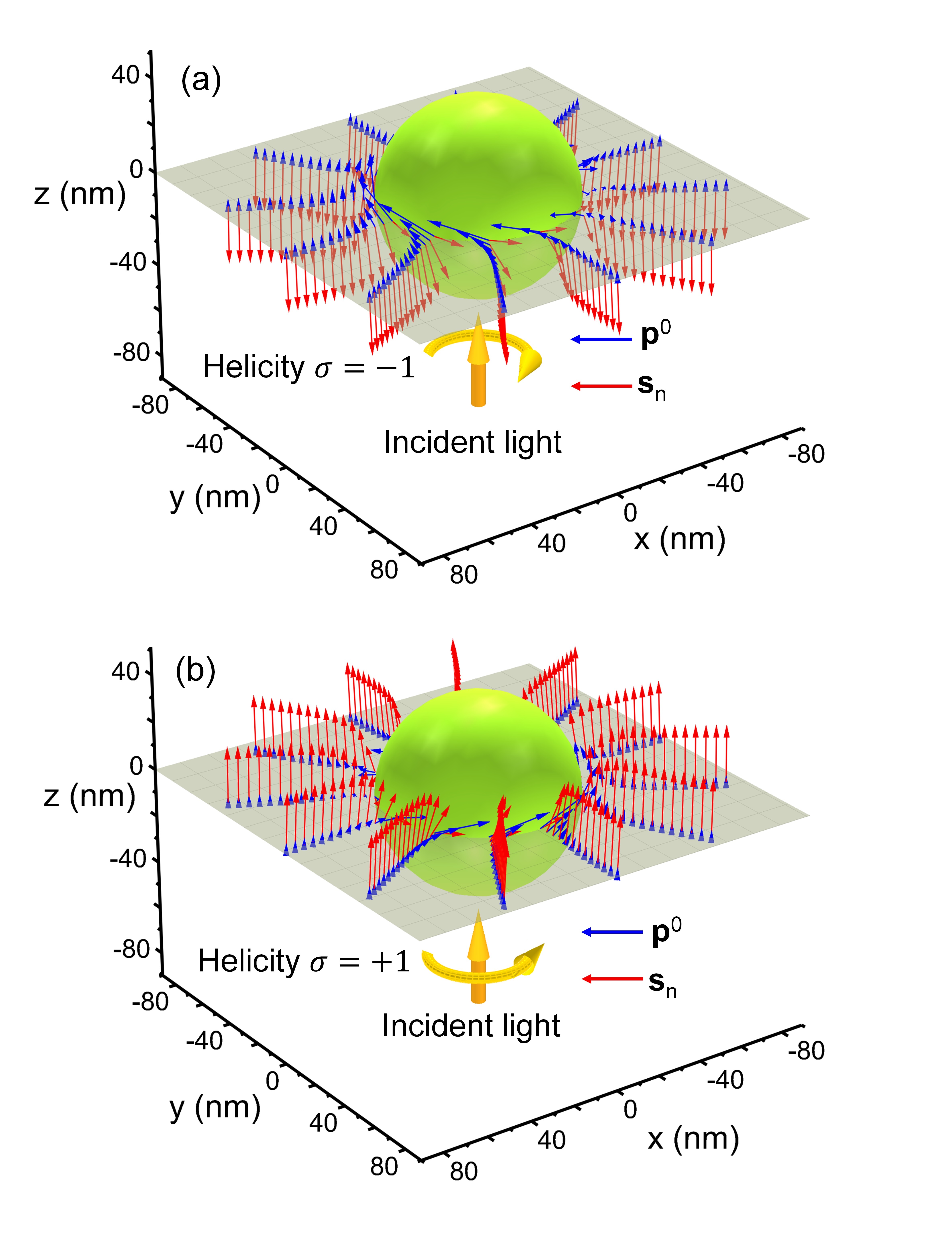}
	\caption{Vectors $\bf p^{\rm o}$ and ${\bf s}_{\rm n}$ displayed on the equatorial plane of the sphere, calculated far from resonance at $\lambda = 735$ nm. (a) for  $\sigma = -1$ incident light  and (b) for  $\sigma = 1$. Note that, in contrast to the results in Fig.~3, $\bf p^{\rm o}$ and ${\bf s}_{\rm n}$ near the surface are almost antiparallel (or almost parallel), depending on the $\sigma$ value of the incident light. This is due to the lower contribution of the scattered field.  
		\label{fig:6}}
\end{figure}

Figure~4 displays the vectors $\bf p^{\rm o}$ and ${\bf s}_{\rm n}$ calculated on a plane 20 nm above the equatorial plane of the sphere, at the plasmonic resonance $\lambda = 531$ nm, for  $\sigma = -1$ incident light. In agreement with the results obtained on the equatorial plane, also in this case the vectors $\bf p^{\rm o}$ and ${\bf s}_{\rm n}$ (close to the particle) remain orthogonal to each other and tangent to the sphere surface.

In order to better understand the impact of LSP resonances, we consider a second excitation wavelength at $735$ nm, quite far from the LSPs near-field peak (see Fig.~1b). Figure~5a shows the orbital momentum enhancement $|{\bf p}^{\rm o}|/|{\bf p_{\rm inc}}|$ and the module of the Poynting vector $|{\bf p}|/|{\bf p_{\rm inc}}|$, normalized with respect to the Poynting vector of the incident field, as a function of the distance $d$ from the nanoparticle surface. The system still gives rise to a significant  ``supermomentum" effect \cite{Bliokh2014}. 
Figure~5b displays the normalized spin density $s_{\rm n}$ for the scattered and total fields as a function of $d$.  For the scattering contribution, the curve decays approximately linearly with increasing $d$, analogously to the resonant case. This indicates that the contribution of the scattering field is largely independent from the resonance condition. A different behaviour characterizes the total field. The spin density of the total field becomes significantly smaller than that of the scattered field for $d \lesssim 20$ nm and on the particle surface it achieves its minimum value. Comparing Figs~2(a,b) and 5(a,b), it is interesting to observe that the two minima in (b) occur approximately in correspondence to the same enhancement $|{\bf p}^{\rm o}|/|{\bf p_{\rm inc}}|$. Hence the results in Fig.~4b share the same explanation  with those in Fig.~2b. Analogous considerations can be done for Figs.~2c and 5c.
Figure~5d displays the angle $\phi$ between $\bf p$ and ${\bf p}^{\rm o}$ as a function of $d$ and calculated for the scattered and total fields. We observe that the angle for the total field is higher with respect to the resonant case and reach its maximum at $d \approx 0$. These differences can be understood observing that the spin of the total field displayed in Fig.~5b shows a rapid variation starting from $d \approx 0$, giving rise to a significant spin momentum.
Figures~6a and 6b display the directions of the canonical momentum and of the spin on the equatorial plane of the particle for the two incident helicities $\sigma = \pm 1$, for 
$\lambda = 735$ nm. In this case, as expected after looking at Fig.~5c, even close to the sphere surface the two vectors are far from being orthogonal. However, interestingly, Fig.~6 shows that also out-of-resonance the incident spin is able to determine the rotation direction of the canonical momentum around the nanosphere.

\subsection*{Linearly-polarized incident field}
\begin{figure}[!ht]
	\centering
	\includegraphics[width = 13.5 cm]{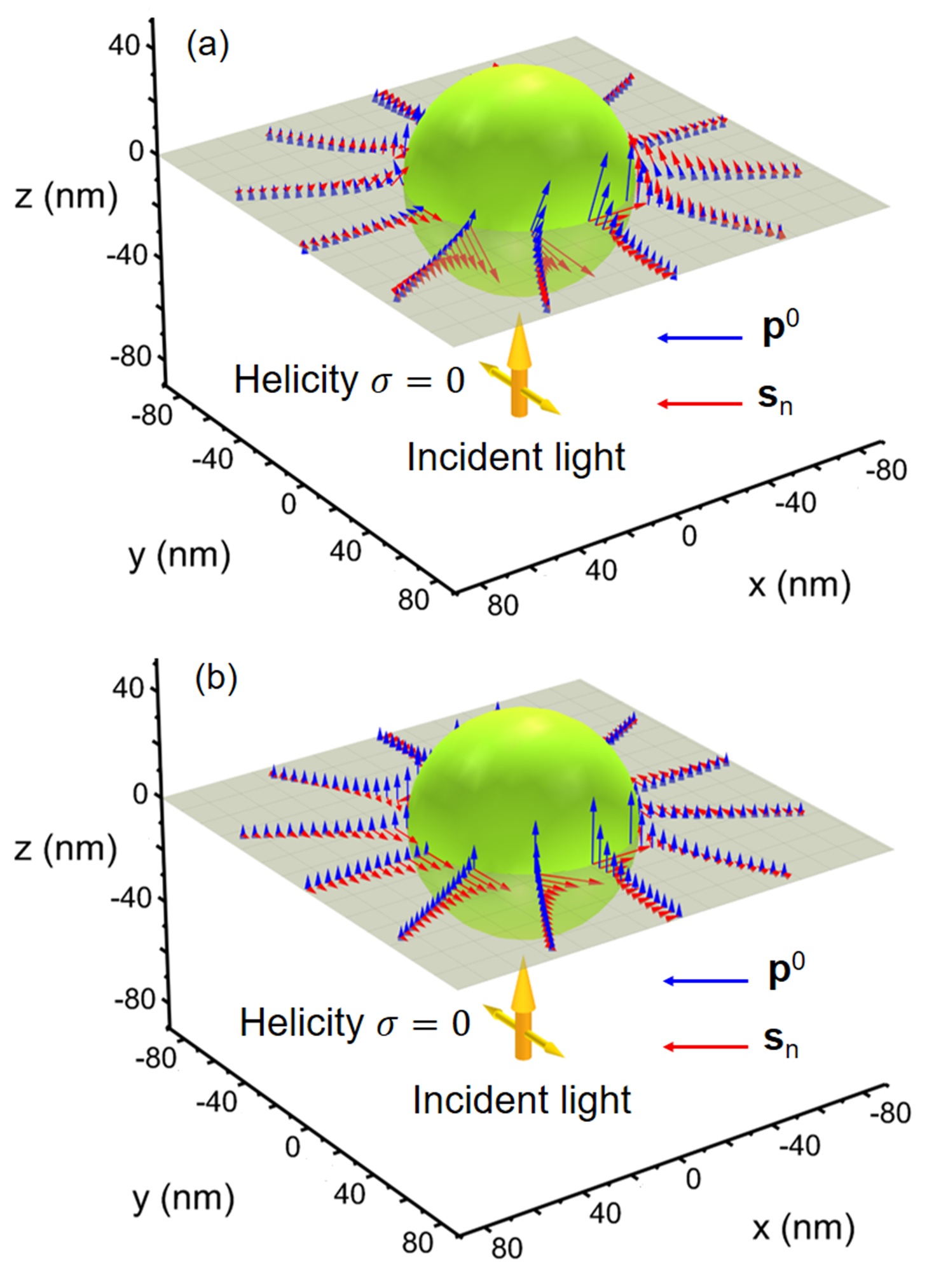}
	\caption{Linearly-polarized incident field: Orbital (or canonical) momentum ${\bf p}^{\rm o}$ and normalized spin ${\bf s}_{\rm n}$ for a gold sphere at the equatorial plane $xy$ for a linearly-polarized ($\sigma=0$) incident field: (a) in the resonance condition ($\lambda=531$ nm), and (b) out of resonance  ($\lambda=735$ nm). 
		\label{fig:7}}
\end{figure}
We now consider a linearly-polarized (zero-spin) plane-wave incident field. Specifically, the incident direction is along the $z$-axis and the polarization direction along the $y$ axis. The system and the incidence direction of the input field are the same as those used for the circular polarization calculations (see Fig.\ 1a). 

Figure~7a displays the logarithm map of the canonical momentum ${\bf p}^{\rm o}$ and of the normalized spin ${\bf s}_{\rm n}$ on the equatorial plane of the nanoparticle. 
The figure shows that around the nanosphere the field acquires spin. This effect origins from the large contribution of evanescent waves in the near-field of a metallic nanoparticle at wavelengths close to a LSP resonance. Indeed, as shown in Eq.~(5), evanescent waves can display a transverse spin even in the absence of an incident spin \cite{Bliokh2014}.
The spin reaches its maximum (${\bf s}_{\rm n} \simeq 0.96$) on the particle surface along the polarization direction ($y$) of the incident field, corresponding also to the direction where the field-enhancement and hence ${\bf p}^{\rm o}$ reach their maximum values. In this direction, the spin vector is along the $x$-axis (orthogonal to both the incident and polarization directions).  We also observe that the angle $\theta$ between  ${\bf s}_{\rm n}$ and ${\bf p}^{\rm o}$ is close to $90^\circ$. However, the canonical momentum  is not exactly along the incident direction. It acquires a small $x$-component. Moving away from the polarization direction, around the nanosphere, both ${\bf s}_{\rm n}$ and ${\bf p}^{\rm o}$
change significantly. The canonical momentum decreases quite rapidly.
Moving clockwise, the spin direction acquires a non-negligible component along the $z$-direction, so that $\theta$ becomes much larger than $90^\circ$. In the $x$-direction, orthogonal to the incident polarizion,
close to the particle surface, the two vectors become antiparallel and both of them reach their minimum (on the surface).

Figure~7b displays the same results reported in Fig.~7a, calculated for an incident wave at  $\lambda = 735$ nm (out of the LSP resonance). We notice that the main difference is that in this case the direction of ${\bf p}^{\rm o}$ almost coincides with the direction of the incident light, in all the points on the equatorial plane.

It is interesting to observe that the  results in Fig.~7 cannot be understood in the dipole (or 
Rayleigh scattering) approximation \cite{Maier2007a}, where it is assumed that the scattered field from a small sphere is well approximated by the field of the dipole moment induced by the incident electromagnetic wave.
For example, according to the dipole approximation, the normalized spin ${\bf s}_{\rm n}$ calculated along  both the $x$ and $y$ directions is zero, in contrast to the results of the full calculations displayed in Fig.~7.
These results have been obtained  by considering that in the vicinity of the surface of the sphere, the scattered field can be expanded in terms of a series of vector spherical Hankel multipole fields \cite{borghese2007scattering}. These fields, solutions to the Maxwell equations and eigenvectors of $L^2$ and $L_z$ as well as of the parity, form a complete set of vectors mutually orthogonal to each other (see, e.g., \cite{borghese2007scattering}).  Unlike what happens in the framework of the Rayleigh scattering approximation, the scattered field, even to the lowest multipole order $L=1$, contains both  radial and transversal parts that significantly affect both the ${\bf s}_{\rm n}$ and ${\bf p}^{\rm o}$ vectors. For example, according to the Rayleigh approximation, the electric field along the direction parallel to the incident field (in the present case the $y$-direction) contains only a radial (longitudinal) contribution, while the exact calculation (even limited to the lowest multipole order $L=1$) contains also a non-negligible  transverse contribution. The presence of both contributions determines a non-zero spin.
\newpage
\section*{Conclusions}

We have investigated the orbital momentum and spin of light and their SOI in the near-field region of a metallic nanoparticle supporting LSP resonances. 
Specifically, we considered circularly or linearly polarized plane waves exciting a gold nanoparticle of radius $a= 40$ nm, considering both the resonant and the non-resonant excitation of the LSP.  All the calculations have been carried out beyond the quasistatic approximation, using the Mie theory implemented within the $T$-matrix formalism \cite{borghese2007scattering}.

We found that the SOI of light in the near-field region gives rise to several interesting features. We summarize the most relevant results:
({\it i}) Due to the strong confinement of the scattered field, the Poynting vector acquires an additional component that depends on the spin. This additional component, so-called spin momentum ${\bf p}^s$, produces a ``supermomentum" effect that causes a strong enhancement of the canonical momentum, which becomes much larger than the Poynting vector.
({\it ii}) The helicity of the circularly polarized incident light is able to control the rotation direction of the canonical momentum ${\bf p}^{\rm o}$ near the surface of the particle (this effect occurs both in the resonant and non-resonant cases).
({\it iii})  In the case of circularly-polarized incident light and for resonant excitations, close to the particle surface the spin is almost opposite to the spin direction of the incident field and is almost orthogonal to the canonical momentum.
({\it iv}) The evanescent waves around the nanoparticle can give rise to significant transverse spin even in the absence of an incident spin. 

Knowledge of the spin and canonical momentum distributions opens way to investigation of optical forces and torques around nanoparticles and nanostructures, which is interesting for experimental studies and applications, since the huge light concentration around metal nanoparticles can give rise to very strong optical forces and torques, even with moderate illumination. The present study can be extended to more complex nanostructures, considering for example metal nano-dimers were very high field-amplification effects in the dimer gap can be obtained at specific wavelengths (see e.g. \cite{nordlander2004plasmon}), and also hybrid nanostructures in the strong \cite{Savasta2010,Ridolfo2011} or ultrastrong \cite{Cacciola2014} light-matter coupling regimes. Moreover, it would be interesting to apply these concepts to  enhanced optical fields and subwavelength-field confinement induced by organic molecules with giant oscillator strength \cite{Gentile2014, Cacciola2015}, and to anisotropic nanoparticles \cite{Bliokh2016}.
Finally, we observe that the analysis developed here, could be useful for the design of optical nano-motors for controlling the motion of even smaller nanoparticles  or molecules (see, e.g., Refs.~\citenum{Bonin2002, Tong2009}).

\section{Methods}

All the calculations presented here were carried out beyond the quasistatic approximation, using the generalized Mie theory \cite{borghese2007scattering,borghese2013superposition}. Near-field and scattering calculations were carried out on a gold nanosphere with radius $a=40$ nm, using a frequency-dependent dielectric permittivity gathered interpolating the experimental data of Ref.~\citenum{Johnson1972}. Calculations have been carried out 
for $\lambda = 531$ nm,  corresponding to the maximum near-field enhancement, and for $\lambda = 735$ nm. At $\lambda = 531$ nm, the interpolated dielectric permittivity of gold is $\varepsilon = -4.616687 + i\,2.3487562$. At $\lambda = 735$ nm, we obtained $\varepsilon = -19.036045 + i\,1.173802$.
In the near-field region, the incident, the internal, and the scattered electromagnetic fields are expanded in vector spherical harmonics (VSH) \cite{borghese2007scattering}. The analytical relations between the incident and scattered multipolar amplitudes are obtained thanks to the linearity of the Maxwell's equations and of the boundary conditions, taking advantage of the expansion of the electromagnetic fields in terms of VSH. From a computational point of view, the numerical calculation of the fields requires the truncation of the multipole expansion of the fields to a suitable order to ensure the numerical stability of the results. Once the fields around the nanoparticle were obtained, we calculated the orbital momentum and the spin density in the near-field region of the nanoparticle by using Eqs.~(\ref{planewave3}) and~(\ref{planewave4}) below. 
In the far-field region, the optical properties of the scatterer have been calculated using the multipolar amplitudes that enter in Mie theory implemented within the T-matrix formalism \cite{borghese2007scattering,borghese2013superposition}. The transition matrix contains all the information on the microphysical properties of the scatterer, being independent from the state of polarization of incidence field and from the incident and observation direction. The elements of the T-matrix define analytically in the far field the optical cross section.

\newpage
\noindent
{\Large \bf Acknowledgements}

FN was partially supported by the RIKEN iTHES Project,
MURI Center for Dynamic Magneto-Optics via the AFOSR Award No. FA9550-14-1-0040,
the Japan Society for the Promotion of Science (KAKENHI),
the IMPACT program of JST,
JSPS-RFBR grant No 17-52-50023,
CREST grant No. JPMJCR1676,
and the Sir John Templeton Foundation.
RS and SS were partially supported by the MPNS
COST Action MP1403 Nanoscale Quantum Optics.

\bibliography{SMLp2}

\end{document}